\documentclass[12pt]{iopart}
\usepackage[latin9]{inputenc}
\usepackage{graphicx}
\usepackage{hyperref}
\usepackage{siunitx}
\usepackage{graphicx}
\usepackage{color}
\usepackage[normalem]{ulem}
\usepackage{bm}  % bold math
\usepackage{cancel}     % in equations
\DeclareSIUnit\gauss{G}

\begin{document}
\title{Subradiance and radiation trapping in cold atoms}
\author{Patrizia Weiss$^1$, Michelle O. Ara\'ujo$^{1,2}$, Robin Kaiser$^1$ \& William Guerin$^1$}
\address{$^1$ Universit\'e C\^ote d'Azur, CNRS, Institut de Physique de Nice, France}
\address{$^2$ CAPES Foundation, Ministry of Education of Brazil, Bras\'ilia, DF 70040-020, Brazil}
\ead{Patrizia.weiss@inphyni.cnrs.fr}
\vspace{10pt}
\begin{indented}
\item[]February 2018
\end{indented}

\begin{abstract}
We experimentally and numerically study the temporal dynamics of light scattered by large clouds of cold atoms after the exciting laser is switched off, in the low intensity (linear optics) regime.
Radiation trapping due to multiple scattering as well as subradiance lead to decay much slower than the single atom fluorescence decay.
These two effects have already been observed separately, but the interplay between them remained to be understood.
Here, we show that with well chosen parameters of the driving field, the two effects can occur at the same time, but follow different scaling behaviors.
The subradiant decay is observed at late time and its rate is independent of the detuning, while the radiation trapping decay is observed at intermediate time and depends on the detuning through the optical depth of the sample. Numerical simulations based on random walk process and coupled-dipole equations support our interpretations. Our study clarifies the different interpretations and physical mechanisms at the origin of slow temporal dynamics of light in cold atoms.
\end{abstract}

\section{Introduction}

Collective effects in light scattering by atomic ensembles have recently been the subject of intense research, both theoretically and experimentally \cite{Guerin2017b,Kupriyanov2017}. %\cite{Goban2015}.
Even in the most simple situation, when the atomic system is driven by a low intensity laser (single-photon or linear-optics regime) and when the atomic cloud has a low density, various phenomena can occur \cite{Ruostekoski1997,Scully2006,Eberly2006,Scully2009}.
For example, steady-state experiments about light diffusion~\cite{Matsko2001,Labeyrie2004}, coherent backscattering~\cite{Labeyrie1999,Bidel2002} and the resonance line shape and shift \cite{Roehlsberger2010,Keaveney2012,Pellegrino2014,Okaba2014,Bromley2016,Jennewein2016,Roof2016,Corman2017} have been performed.
Several recent experiments also studied the temporal dynamics of the light scattered by cold atoms at the switch off of the driving field.
A decay faster than the natural decay rate $\Gamma$ has been observed at short time, a signature of superradiance~\cite{Roof2016, Araujo2016}.
A decay rate much slower than $\Gamma$ has also been detected at later time, a direct observation of subradiance~\cite{Guerin2016}.
It has been shown experimentally that the subradiant decay rate depends on the resonant optical depth $b_0$, independently of the detuning $\Delta = \omega-\omega_0$ from the atomic resonance $\omega_0$, which has been confirmed by numerical simulations~\cite{Guerin2016,Bienaime2012,Araujo2018}.

Interestingly, a slow decay can also be interpreted completely differently.
Indeed, near resonance, when the actual optical depth $b(\Delta) \propto b_0/(1+4\Delta^2/\Gamma^2)$ is large, light undergoes multiple scattering.
This leads to a slowed transport velocity inside the diffusive medium~\cite{Lagendijk1996} and ultimately to a slow decay when the incident light is switched off.
This effect, called radiation trapping~\cite{Kenty1932,Holstein1947,Molisch1998}, has also been studied in cold atoms~\cite{Fioretti1998,Labeyrie2003,Labeyrie2005,Balik2005,Pierrat2009,Balik2013}.
In particular, it has been shown that, at low enough temperature, the dynamics depends on the detuning only through the optical depth $b(\Delta)$, because this parameter controls the distribution of the number of scattering events that light undergoes before escaping, the average time between scattering events being remarkably independent of the detuning~\cite{Labeyrie2003}.

Radiation suppression can be obtained by different physical mechanisms, as already pointed out by Cummings \cite{Cummings1986} who noted that interference-based radiation suppression is "much more exotic and unexpected than the ordinary radiation trapping", which can be explained by photon rescattering.
As the different scalings [$b_0$ \textit{vs} $b^2(\Delta)$] show, these two effects are not two different interpretations of the same phenomena, but are really due to two different physical mechanisms.
This difference does not appear when one studies the eigenvalues of the effective Hamiltonian describing the atoms interacting through the shared excitation~\cite{Rusek1996,Rusek2000,Pinheiro2004,Skipetrov2014,Bellando2014}, all long-lived collective atomic modes being often called ``subradiant'', although differences in the shape of the eigenmodes have been discussed as a possible way to distinguish between modes associated to subradiance and to radiation trapping ~\cite{Guerin2017a}.

In this article, we experimentally study these two effects, showing in particular that, with well chosen parameters, both occur simultaneously.
We find that when the atomic sample is driven by a plane wave, as in ref.~\cite{Guerin2016}, subradiance is observed and radiation trapping is not clearly visible, even on resonance, mainly because the signal is dominated by single scattering occurring on the edges of the sample.
The situation is different with an exciting beam much smaller than the cloud, as in ref.~\cite{Labeyrie2003}, because single scattering is strongly reduced if light is detected near the forward direction.
In this paper we show that with reduced single scattering near resonance, a slow decay due to radiation trapping is visible at intermediate time and, at later time, an even slower decay appears due to subradiance.
Although at zero temperature and for large enough optical depth, radiation trapping could be slower than subradiance and dominate even at late time, the frequency redistribution due to Doppler broadening strongly reduces the number of scattering events that light can undergo before escaping, and we find that, at $T \sim $ \SI{100}{\micro \kelvin}, subradiant decay always dominates at late time.

The paper is organized as follows.
In the next section we present the experimental setup and in the following the observation of subradiance for an excitation with a plane wave.
In section \ref{sec:both} we present the data acquired with a narrow driving beam, showing the simultaneous observation of subradiance and radiation trapping.
We study in detail how the corresponding decay times scale with the parameters.
In section \ref{sec:num} we present numerical simulations which support our interpretations.
In particular, the comparison between the simulations based on the coupled-dipole equations and on a random walk model performed at $T=0$ allows us to discuss the physics in an ideal case.
Moreover, the simulations based on the random walk model including the effect of the temperature, laser spectrum and beam size are in fair agreement with our experimental data on radiation trapping.
We finally conclude in section \ref{sec:summary}.

\section{Experimental setup}\label{sec:setup}

In the experiment, we prepare a cloud of cold rubidium-87 atoms in a magneto-optical trap (MOT), which is loaded during \SI{60}{\milli \second} from the background vapor in the glass chamber.
For further increase of the optical depth a compressed MOT stage follows for \SI{30}{\milli \second}, which additionally leads to a cleaner shape of the cloud (close to a Gaussian density distribution) and a reduced temperature.
We obtain an ensemble of $N\approx 2.5\times10^9$ atoms at a temperature $T\approx$ \SI{100}{\micro \kelvin}.
After switching off all MOT beams as well as the magnetic fields, the cloud is allowed to expand ballistically for a duration of \SI{3}{\milli \second}, during which the atoms are optically pumped to the upper hyperfine ground state $F=2$.

\begin{figure}
\centerline{\includegraphics{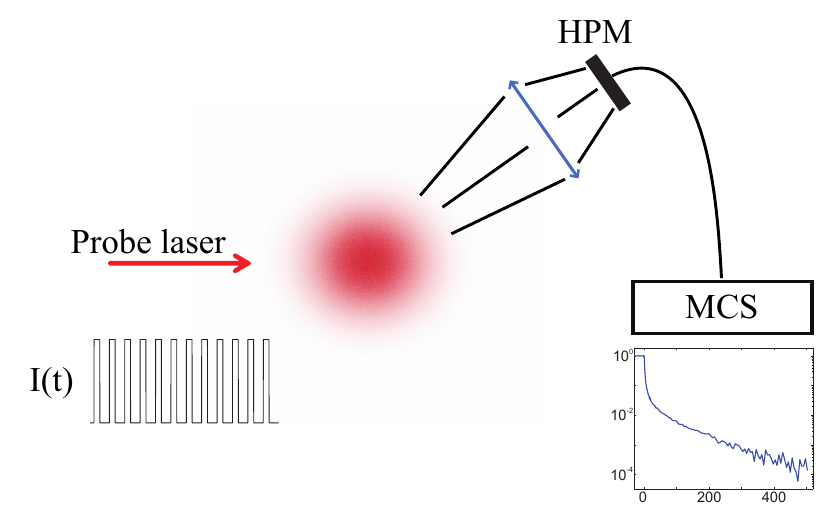}}
\caption{The experimental set-up consists of a cold cloud of $^{87}$Rb atoms, prepared in a MOT. This cloud is excited with a probe beam of variable size. After a fast switch-off the scattered light is collected under an angle of \SI{35}{\degree} with a hybrid photo multiplier (HPM). The signal is recorded with a multichannel scaler (MCS). During the free expansion of the cloud a series of 12 pulses is applied, during which the optical depth evolves.}
\label{fig:setup}
\end{figure}

After this preparation stage the typical peak density is $\rho_0 \sim 10^{11}$ \SI{}{\centi \meter ^{-3}} and the rms size is $R \approx$\SI{1}{\milli \meter}.
To weakly excite the cloud a series of $12$ pulses are applied, each of them with a duration of \SI{10}{\micro \second} and a separation of \SI{1}{\milli \second}.
The probe beam is generated by a commercial external-cavity diode laser with a linewidth of $FWHM=$ \SI{500}{\kilo \hertz} \cite{footnote1}.
The probe laser has a linear polarization and a normalized detuning to the atomic resonance of $\delta = \left(\omega - \omega_0\right)/\Gamma$, where $\omega$ is the laser frequency, $\omega_0$ the atomic transition frequency of the $F=2 \rightarrow F'=3$ transition and $\Gamma/2\pi=$ \SI{6,07}{\mega \hertz} is the natural linewidth.
We ensure that we stay in the weak excitation limit by adapting the probe intensity to the detuning $\delta$, such that the saturation parameter 
\begin{equation}
s(\delta)=g \frac{I/I_\text{sat}}{1+4\delta^2}\, 
\end{equation}
remains small, with $I_\text{sat} = 1.6$ \SI{}{\milli\watt/ \centi\meter^{2}} and $g=7/15$ the degeneracy factor of the transition for equipopulated Zeeman states.
The dynamic range for the light detection is mainly determined by the extinction ratio of the probe, which is achieved to a level of $10^{-4}$ by using two acousto-optical modulators in series and being satisfactory faster ($t_\text{switch} \approx$ \SI{15}{\nano \second}) than the natural lifetime of the excited state, $\tau_\text{at}=\Gamma^{-1}=$ \SI{26}{\nano \second}.
Due to the free expansion of the cloud during the pulse series, the optical depth changes for every pulse.
After the pulse series the MOT is turned on again and most of the atoms are recaptured.
This leads to a total cycle duration below \SI{150}{\milli \second} and allows averaging over a large number of cycles ($\sim 500\: 000$) for each measurement.
As sketched in figure \ref{fig:setup} the scattered light is collected via a two-inch lens under an angle of \SI{35}{\degree} and collected by a hybrid photomultiplier (HPM, ref. R10467U-50 from Hamamatsu).
The signal is recorded via a multichannel scaler (MCS) with a time resolution of \SI{1,6}{\nano \second} while averaging over the cycles.

The optical depth during the pulse series is calibrated afterwards via absorption imaging~\cite{footnote_SuppMat}.
In the following we will note $b_0$ the optical depth of the cloud on resonance assuming the Clebsch-Gordan coefficient of the transition is unity, which corresponds for a Gaussian cloud to $b_0=3N/(kR)^2$ with $N$ the atom number and $R$ the rms radius. The actual detuning-dependent optical depth is then given by
\begin{equation}
b(\delta) = g \frac{b_0}{1+4\delta^2} \, ,
\end{equation}
including the degeneracy factor $g=7/15$ of the probed transition.

\section{Observation of subradiance} \label{sec:sub}

The direct observation of subradiance for a large number of atoms $N$ was accomplished in~\cite{Guerin2016}.
We present here similar measurements to confirm the results with the upgraded set-up~\cite{footnote1}, as well as to serve as a reference for the following measurements.
\begin{figure}
\centerline{\includegraphics[width=\columnwidth]{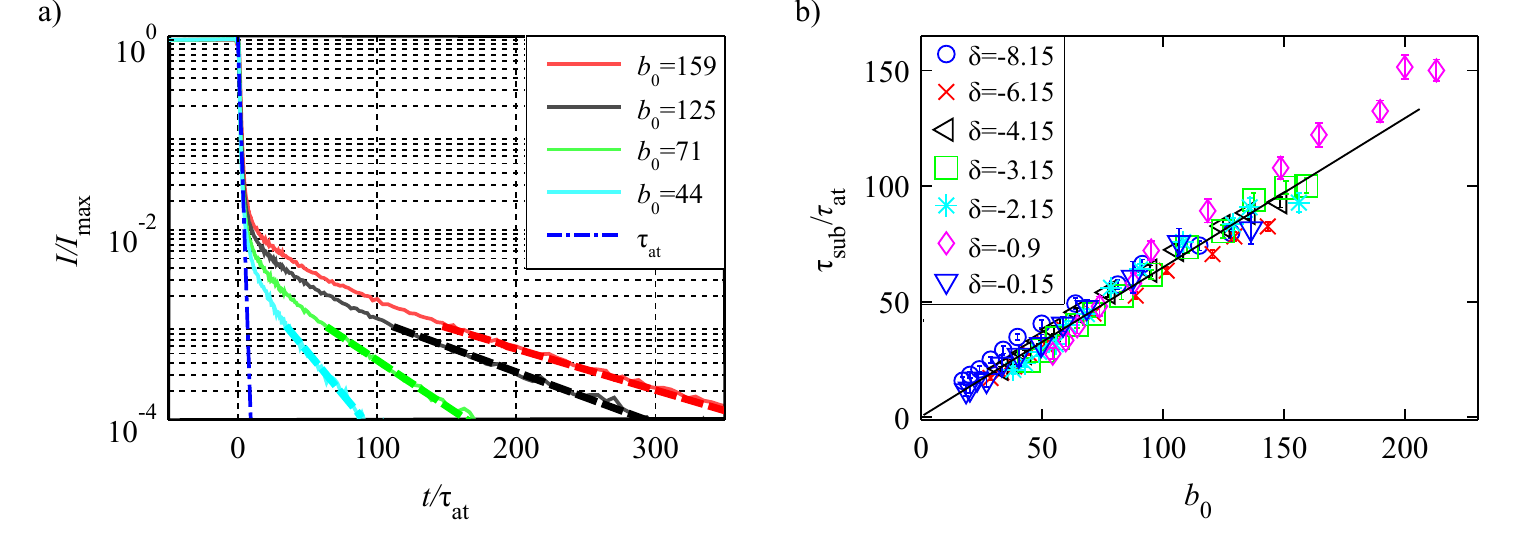}}
\caption{(a) Experimental decay curves for different $b_0$, measured with a normalized detuning of $\delta=-3.15$. All curves are normalized to the level right at the switch off of the probe beam. For comparison, the theoretical single atom decay $\tau_\text{at}$ is also shown (dash-dotted line). The slowest decay time $\tau_\text{sub}$ is determined by an exponential fit (dashed lines) at late time. (b) Measured subradiance decay times $\tau_\text{sub}/\tau_\text{at}$ as a function of the on-resonance optical depth $b_0$. All measured points collapse on a single line, independent of the detuning. The linear scaling of $\tau_\text{sub}$ with $b_0$ is stressed by the linear fit (solid line).}
\label{fig:sub_large}
\end{figure}

In this section, we use a driving beam which is much larger in diameter than the atomic cloud, with a waist ($1/e^2$ radius) $w=5.7$ \SI{}{\milli\meter}, creating a homogenous excitation of the cloud.
The saturation parameter is set to $s(\delta)\approx 0.02$.
In figure \ref{fig:sub_large}(a) an example of a data set acquired with a detuning of $\delta=-3.15$ is shown.
Four decay curves are plotted, corresponding to different pulses and thus to different values for $b_0$.
After an initial fast decay down to an amplitude of $\sim 10^{-2}$ relative to the steady-state level (before switch-off), a very slow decay is well visible, with a time constant that clearly changes with $b_0$.
To characterize this time constant, we choose to fit the experimental decay curve by a single decaying exponential in a range defined as one decade above the noise floor. This procedure thus corresponds to the longest visible decay time.

We performed a series of measurements for different detunings $\delta$.
The measured time constants $\tau_\text{sub}$, in unit of the single atom decay time $\tau_\text{at}$, are shown in figure~\ref{fig:sub_large}(b) as a function of the on-resonant optical depth $b_0$ for the different detunings.
All points collapse on a single curve, well fitted by a single line with $\tau_\text{sub}/\tau_\text{at} \approx 1+0.65 \:b_0$.
This demonstrates that this longest decay time is independent of the detuning and scales linearly with $b_0$, in perfect agreement with the expectations for subradiance~\cite{Guerin2017b,Guerin2016,Bienaime2012,Araujo2018}.

\section{Simultaneous observation of radiation trapping and subradiance} \label{sec:both}

As the data of figure~\ref{fig:sub_large}(b) show, the decay rate at long time is independent of the detuning, even close to resonance.
This fact might come surprising, since close to resonance, the actual optical depth $b(\delta)$ is large, which induces attenuation of the driving beam inside the sample and multiple scattering.
It has been shown in previous studies that this indeed leads to a suppression of some cooperative effects close to resonance, i.e. the fast decaying modes of superradiance~\cite{Araujo2016,Guerin2017a}.
Nevertheless, the slow-decaying modes remain visible and are even enhanced on resonance~\cite{Guerin2016,Guerin2017a}.
This raises the question of the interpretation of these slow-decaying modes near resonance: subradiance or radiation trapping due to multiple scattering?

\subsection{Classical description of radiation trapping}

To describe multiple scattering of light, the basic quantity is the mean-free path $\ell_\text{sc} = 1/(\rho \sigma_\text{sc})$, where $\rho$ is the density of scatterers and $\sigma_\text{sc}$ their scattering cross-section.
We suppose here that the scattering diagram is isotropic, which is a good approximation for multi-level Rb atoms, where all Zeeman-sublevels of the $F=2$ ground state are equally populated \cite{Greene1982}.

\begin{figure}
\centerline{\includegraphics{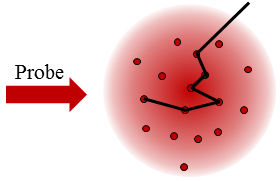}}
\caption{Classical picture of radiation trapping. A narrow probe beam near resonance is sent onto the atomic sample, considered as an ensemble of point-like scatterers. Light undergoes multiple scattering events inside the vapor before escaping.}
\label{fig:RT}
\end{figure}

In a scattering medium of size much larger than the mean-free path (large optical depth), light is scattered many times before escaping (figure~\ref{fig:RT}).
In this case, many observables can be very well described by a diffusion equation for the electromagnetic energy density, at the condition to perform an average over the disorder configurations~\cite{Johnson2003}.
In three dimensions the spatial diffusion coefficient reads
\begin{equation}\label{eq:D}
D = \frac{v_E \ell_\text{sc}}{3} = \frac{\ell_\text{sc}^2}{3\tau_\text{tr}} \, ,
\end{equation}
where $v_E = \ell_\text{sc}/\tau_\text{tr}$ is the energy transport velocity inside the medium and $\tau_\text{tr}$ the transport time~\cite{Rossum1999}.
The transport time is the sum of the group delay between two scattering events and the delay associated with the elastic scattering process, called Wigner's delay time $\tau_\text{W}$~\cite{Lagendijk1996,Labeyrie2003}:
\begin{equation}\label{eq:t_tr}
\tau_\text{tr} = \tau_\text{W} + \frac{\ell_\text{sc}}{v_\text{g}} \, ,
\end{equation}
where $v_\text{g}$ is the group velocity.
For near-resonant light, a remarkable property of cold atomic vapor is that $\tau_\text{tr} = \tau_\text{at}$, the lifetime of the excited state, independently of the detuning~\cite{Lagendijk1996,Labeyrie2003} (see \ref{sec:A_tau} for discussion and full derivation of this property).

As a consequence, the temporal dynamics of the diffuse light is mainly governed by the number of scattering events $\left\langle N_\text{sc}\right\rangle$ that light undergoes before escaping the atomic cloud.
%After some time of illumination, a sudden switch-off of the exciting laser leads to a slow decrease of the scattered light due to multiple scattering. This `imprisonment of radiation'~\cite{Holstein1947}, or `radiation trapping'~\cite{Molisch1998}, has been studied in cold atoms~\cite{Fioretti1998,Labeyrie2003}, taking also into account subtle effects like the frequency redistribution induced by Doppler broadening~\cite{Labeyrie2005,Pierrat2009} or the multilevel structure~\cite{Baudouin2013a}. Neglecting those effects, one can easily find the scaling of the radiation trapping time with the optical depth.
This number can be evaluated from hand-waving arguments based on a diffusion process. In 3D, the energy density spreads as $\left\langle r^2 \right\rangle = 6 D t$.
Then the average number of scattering events for escaping photons is the ratio between the time spent in the system and the scattering time $\tau_\text{at}$,
\begin{equation}
\left\langle N_\text{sc}\right\rangle = \frac{t}{\tau_\text{at}} = \frac{\left\langle r^2\right\rangle}{6 D \tau_\text{at}}.
\end{equation}
The radiation can escape the system when $\sqrt{\langle r^2 \rangle} \sim R = b \ell_\text{sc}/2$. Using $D=\ell_\text{sc}^2/(3\tau_\text{at})$ leads to $\left\langle N_\text{sc}\right\rangle \sim b^2/8$.
In the diffusive regime (large $b$), radiation trapping times are thus expected to scale as $b^2$, with a precise numerical prefactor that depends on the geometry of the medium~\cite{Labeyrie2003,Cao2003b}.

Since radiation trapping scales as $b^2$ and subradiance as $b_0$, one can expect that for large enough $b$, radiation trapping leads to a slower decay than subradiance and dominates the long-time dynamics.
As we will see in section~\ref{sec:numT0}, this is indeed what numerical simulations performed at zero temperature show.

However, frequency redistribution due to Doppler broadening breaks the $b^2$ scaling. Indeed, at each scattering event, light is Doppler shifted by only a small amount, but at large optical thickness the number of scattering events becomes large and a part of the light eventually gets out of resonance.
This mechanism thus limits the number of scattering events, and consequently the characteristic time of radiation trapping \cite{Labeyrie2003, Labeyrie2005}, which scales almost linearly with $b$ \cite{Pierrat2009}.
There is however, to our knowledge, no analytical description of radiation trapping in this regime and one has to use numerical simulations including the frequency redistribution to describe the decay dynamics.
Such simulations will be discussed in section~\ref{sec:num_T}.

\subsection{Impact of the probe beam size}

Beside the time scale of radiation trapping, the relative amplitude of the slow-decaying part of the signal is of paramount importance to be able to observe radiation trapping.
This is largely related to the relative proportion of multiply-scattered light in the detected signal, which is itself related to the geometry of the experiment, especially the size of the exciting beam, the cloud shape and detection direction.

\begin{figure}
\centerline{\includegraphics{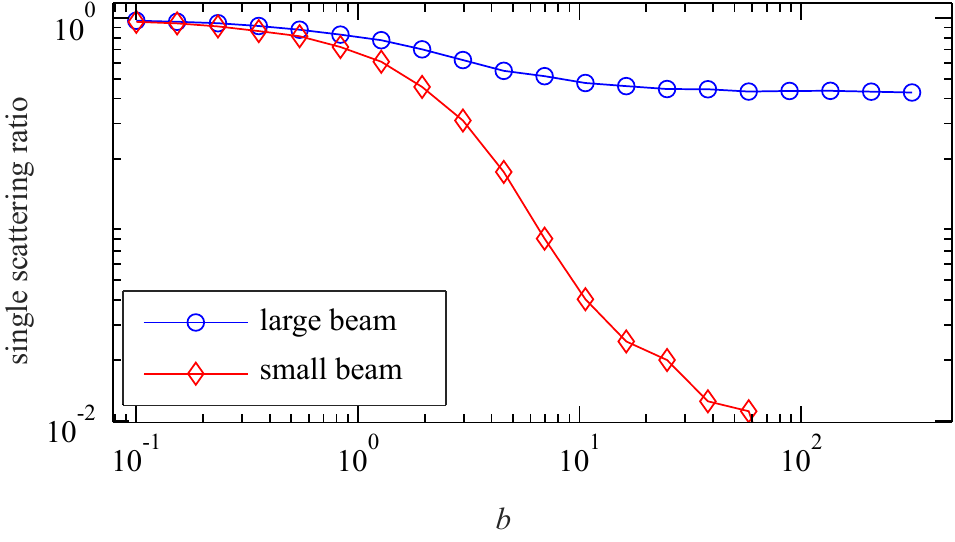}}
\caption{Numerical simulations for the proportion of photons having undergone only one scattering event before escaping in the detection direction, at $\theta = 35^\circ \pm 10^\circ$ from the incident direction, as a function of the optical depth $b$, obtained from random walk simulations. Blue circles are for an illumination with a plane wave and red diamonds for an infinitely narrow beam centered on the Gaussian cloud. For large $b$ single scattering is suppressed with a very narrow beam but remains quite high with a plane wave.}
\label{fig:scat}
\end{figure}

We illustrate this by showing in figure~\ref{fig:scat} the proportion of photons having undergone only one scattering event before escaping the sample in the detection direction, for excitation with a plane wave and with a beam sufficiently smaller than the cloud.
It shows that for large optical depth, single scattering is suppressed with a very narrow beam, as is intuitively expected, and so the detected signal is almost exclusively due to multiply-scattered light.
This is very different for an illuminating beam larger than the cloud, like a plane wave, because a non-negligible proportion of the incoming light will probe the edges of the atomic cloud, where the optical depth is much lower, and slowly tends to zero with a Gaussian cloud.
Therefore there is always a large proportion of single and low-order scattering, even for very large optical depth $b$ (defined for light crossing the cloud along its center).

For the subradiance measurement presented in ref.~\cite{Guerin2016} and in section~\ref{fig:sub_large}, the probe beam is much larger than the atomic cloud, which leads to a dominant contribution of single and low-order scattering, even on resonance.
The slow decay that could be due to radiation trapping has thus a reduced relative amplitude, and subradiance dominates.

In order to study radiation trapping, it is thus necessary to use a driving beam significantly smaller than the size of the atomic sample, as in ref.~\cite{Labeyrie2003}.
We will use in the following a beam with a waist $w =$ \SI{200}{\micro \meter}, well below the radius of the atomic cloud.

The strong reduction of the beam size comes along with several experimental difficulties.
First, the intensity has to remain low enough in order to keep the saturation parameter still small, which for a narrow beam size corresponds to very low power, and thus a reduced detected signal. Second, because of multiple scattering, the amount of scattered light near the forward direction \emph{decreases} when the optical depth increases~\cite{Labeyrie2004}, much more strongly than with a plane wave where light is transmitted near the edges.
As a consequence we were not able to acquire data with a sufficient dynamics for detunings very close to resonance, and the dynamics of the recorded decay curves with a narrow beam is not as good as those recorded with a plane wave (more than 4 decades in figure~\ref{fig:sub_large}).
Nevertheless, we were able to obtain clear signatures of radiation trapping and subradiance, as detailed in the following.

\subsection{Measurements and data analysis}

\begin{figure}
\centerline{\includegraphics{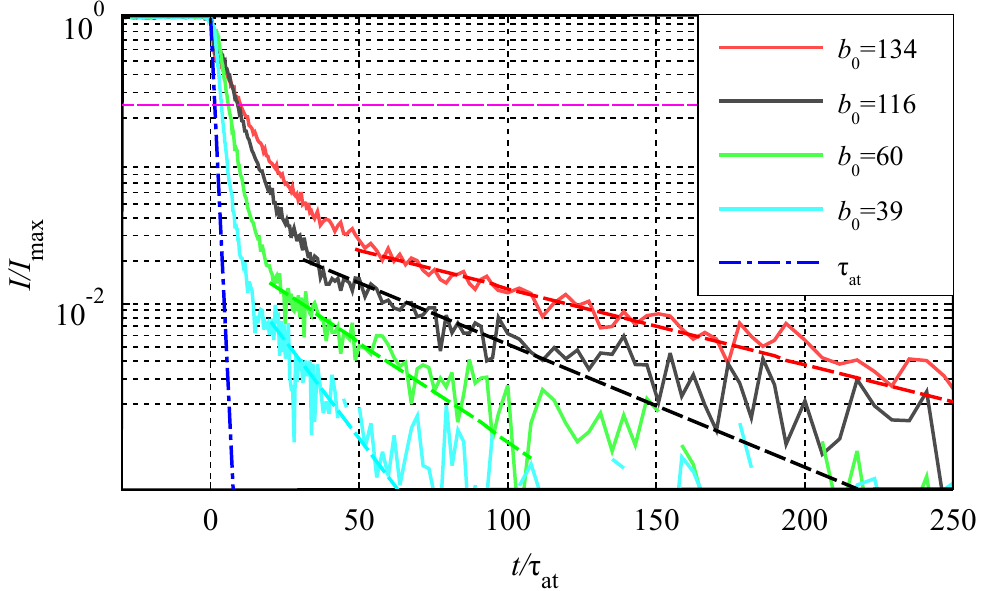}}
\caption{Experimental decay curves for different values of $b_0$ at a fixed detuning $\delta=-0.9$ and a narrow probe beam. Additionally to a very slow decay at late time, similar to the one observed with a plane wave (figure~\ref{fig:sub_large}), another slow decay appears at intermediate time. This intermediate decay is also slower than the natural decay time $\tau_\text{at}$ (dash-dotted line). The fit result obtained by a single exponential for the slowest decay is shown with the dashed lines, and the level used to characterize the intermediate decay time is shown as a horizontal magenta dashed line.}
\label{fig:decays_small}
\end{figure}

The experimental setup and procedure is the same as described in section \ref{sec:setup}, except the size of the probe beam, which now has a waist of $w =$ \SI{200}{\micro \meter}.
Measurements with this narrow beam are shown in figure~\ref{fig:decays_small}.
The decay curves are averaged over 600 000 cycles and the different values for the optical depth are again due to the free expansion of the cloud during the pulse series.
The curves are recorded for a detuning of $\delta=-0.9$, which is close enough to resonance to be in the multiple scattering regime ($b(\delta)\gg1$).
At long time, we clearly observe a very slow decay similar to the subradiant decay observed with a plane wave (figure~\ref{fig:sub_large}).
However, the decay at short and intermediate time is now much slower than in the plane wave case.
The two parts of the decay curves evolve both with the optical depth.

In order to interpret these curves and identify the physical mechanisms at the origin of the two slow decays, we have performed systematic measurements for several $b_0$ and $\delta$. We have kept the saturation parameter lower than $0.4$ for all data and the lowest count rate in steady-state was $6\times 10^5$ counts per seconds.

In order to characterize those decays by simple numbers, we have used the following procedure.
For the late-time decay we use a single exponential fit and we keep the same fitting range as for the previous measurements with a plane wave, i.e. one decade above the noise floor.
The characterization of the intermediate decay is less straightforward since it is clearly not a single exponential decay.
We have chosen to measure the time at which the normalized intensity reaches $e^{-1}=36.8\%$ as an effective decay time.
This level seemed a good trade-off between waiting long enough such that the fastest modes have decayed and not too long not to enter the late-time decay.
A reliable determination of this time has to take into account the non-negligible amount of detected light which does not come from the cold atoms but from the scattering off the glass windows and the background hot vapor, such that the 36.8\% level is always defined respective to the steady-state level of the light scattered by the cold atoms.
The corresponding level is shown in figure~\ref{fig:decays_small} as a dashed horizontal line.

\begin{figure}
\centerline{\includegraphics[width=\columnwidth]{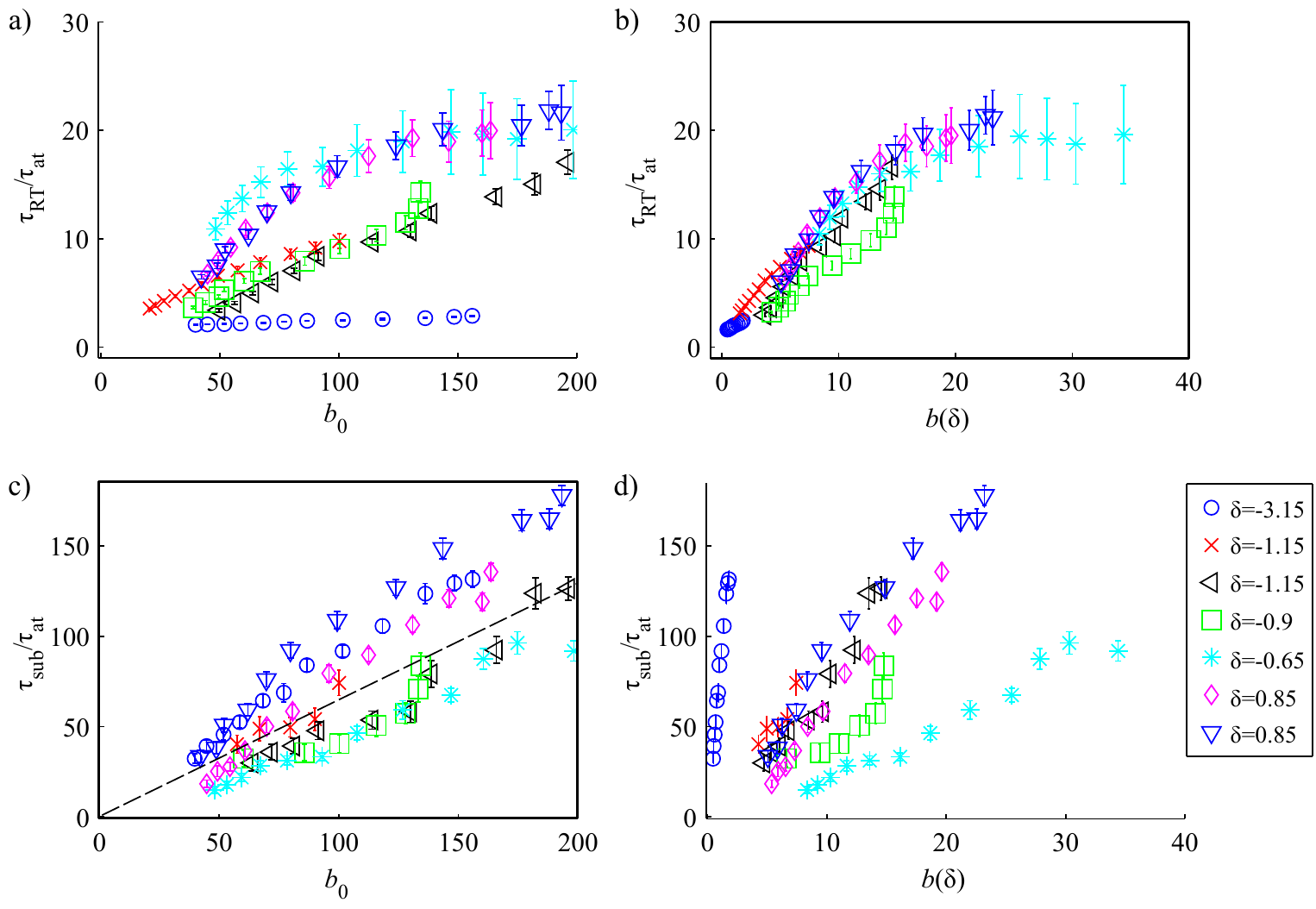}}
\caption{Systematic experimental study of the decay of the scattered light when the atomic sample is illuminated by a narrow beam. (a) and (b) Intermediate decay time $\tau_\text{RT}$ plotted as a function of $b_0$ and $b(\delta)$. (c) and (d) Late-time decay $\tau_\text{sub}$ as a function of $b_0$ and $b(\delta)$. The relevant scalings appear in panels (b) and (c). In the latter the subradiance trend measured with a plane wave (figure~\ref{fig:sub_large}) is shown as a dashed line.}
\label{fig:tau_small}
\end{figure}

The results of the measured decay times for several $b_0$ and $\delta$ are shown in figure~\ref{fig:tau_small}.
In the first row [panels (a) and (b)], we plot the effective decay time characterizing the intermediate decay, noted $\tau_\text{RT}$, and in the second row [panels (c) and (d)] we plot the slowest decay time, noted $\tau_\text{sub}$.
Moreover, in order to identify the relevant scaling parameter for each decay time, we plot them as a function of $b_0$ [left panels (a) and (c)] and $b(\delta)$ [right panels (b) and (d)].
One can see that $b_0$ is not the right parameter governing the intermediate decay (figure \ref{fig:tau_small}(a)) and $b(\delta)$ is not the right parameter governing the long-time decay (figure \ref{fig:tau_small}(d)).
The relevant scaling are those of figures~\ref{fig:tau_small}(b,c), highlighted by thick mirrored axes.

The measured values $\tau_\text{RT}$ for the intermediate decay plotted as a function of $b(\delta)$ all collapse quite well on a single curve, showing that the optical thickness governs this decay.
We therefore associate this to radiation trapping. Note that $\tau_\text{RT}$ does not scale as $b^2$, which is partly due to the Doppler-induced frequency redistribution, as already explained, and also partly due to our empirical definition of $\tau_\text{RT}$, which does not correspond to the lifetime of the longest-lived diffusive mode.
The data is in fair agreement with the random walk simulations shown in section~\ref{sec:num_T}, which demonstrates that classical multiple scattering is a sufficient ingredient to explain this part of the decay curve.
However we note that the scaling with $b(\delta)$ has been obtained using an empirical frequency shift of $-0.15\Gamma \approx 0.9\,$ MHz for the probe detuning, which might be due to calibration errors or spurious magnetic fields.
All data are presented with this shifted detuning.

The measured values $\tau_\text{sub}$ for the slowest decay time plotted as a function of $b_0$ are scattered around the trend of the subradiance decay measured with the plane wave (figure \ref{fig:sub_large}), shown by the dashed line.
We do not observe any significant systematic effect with the detuning. The higher level of the noise floor compared to the plane wave data, due to the reduced probe power, explains the spreading of the data, but the trend shows unambiguously that this decay is similar to the one observed with the plane wave, and can thus be attributed to subradiance.
As a consequence, we can conclude that with these parameters, in particular the temperature $T\approx$ \SI{100}{\micro \kelvin}, the late-time decay is dominated by subradiance, even with a narrow exciting beam, at least up to $b \sim 35$, which is the maximum we have been able to study in our experiment.

\section{Numerical simulations}\label{sec:num}

In order to provide further evidence of our interpretation to distinguish radiation trapping from subradiance, we now turn to numerical simulations.
Numerical simulations allow us to discuss the physics of subradiance and radiation trapping in an idealized scenario, for example at zero temperature.
It also allows comparing the data to a model including a number of experimental imperfections.

\subsection{Description of the models}

We use two very different models in the following: coupled-dipole (CD) equations and random walk simulations (RW).

The coupled-dipole model has been widely used in the last years in the context of single-photon superradiance and subradiance~\cite{Scully2006,Roof2016,Araujo2016,Guerin2016,Bienaime2012,Scully2009a,Courteille2010,Svidzinsky2010,Bienaime2011,Bienaime2013}.
It considers $N$ two-level atoms at random positions $\bm{r}_i$ driven by an incident laser (Rabi frequency $\Omega(\bm{r})$, detuning $\Delta$).
Restricting the Hilbert space to the subspace spanned by the ground state of the atoms $|G\rangle = |g \cdots g \rangle$ and the singly-excited states $|i\rangle = |g \cdots e_i \cdots g\rangle$ and tracing over the photon degrees of freedom, one obtains an effective Hamiltonian describing the time evolution of the atomic wave function $| \psi(t) \rangle$,
\begin{equation}
| \psi(t) \rangle = \alpha(t) | G \rangle +  \sum\limits_{i=1}^N \beta_{i}(t)| i \rangle \; . \label{eq:psi}
\end{equation}
Considering the low intensity limit, when atoms are mainly in their ground states, i.e. $\alpha \simeq 1$, the problem amounts to determine the amplitudes $\beta_i$, which are then given by the linear system of coupled equations
\begin{equation}
\dot{\beta}_i = \left( i\Delta-\frac{\Gamma}{2} \right)\beta_i -\frac{i\Omega_i}{2} + \frac{i\Gamma}{2} \sum_{i \neq j} V_{ij}\beta_j \; .
\label{eq:betas}
\end{equation}
These equations are the same as those describing $N$ classical dipoles driven by an oscillating electric field~\cite{Svidzinsky2010}.
The first term on the left hand side corresponds to the natural evolution of independent dipoles, the second one to the driving by the external laser, the last term corresponds to the dipole-dipole interaction and is responsible for all collective effects.  In the scalar model for light, which neglects polarization effects and near-field terms in the dipole-dipole interaction, it reads
\begin{equation}
V_{ij} = \frac{e^{ik_0r_{ij}}}{k_0r_{ij}} \, , \mathrm{with} \; r_{ij} = |\bm{r}_i - \bm{r}_j| \; , \label{eq:Vij}
\end{equation}
where $k_0=\omega_0/c$ is the wavevector associated to the transition.
Neglecting the near field terms of the dipole-dipole interaction is a good approximation for dilute clouds, i.e. when the typical distance between atoms is much larger than the wavelength, which is the case in the experiment.
The impact of the polarization of light on subradiance, as well as the Zeeman structure of the atoms, is still an open question and has been the subject of several recent theoretical works \cite{Lee2016, Hebenstreit2017, Sutherland2017}
From the computed values of $\beta_i$, we can derive the intensity of the light radiated by the cloud as a function of time and of the angle~\cite{Bienaime2011}.
Technical details on the simulations can be found in~\cite{Araujo2018}.

The second model is a random-walk model, where the atoms are treated as classical scatterers and photons as particles, neglecting wave aspects.
Photons are sent one by one by randomly drawing their initial transverse position according to the exciting laser profile and their initial detuning according to the laser spectrum.
The number of scattering events until the photon escapes the medium, as well as its escape direction, are computed from a stochastic algorithm based on the mean-free path~\cite{Molisch1998}.
By repeating this with many photons, we can build the distribution of the number of scattering events per photon for a given detection direction.
By converting the number of scattering to a time using the transport time $\tau_\text{at}$ (see~\ref{sec:A_tau}) and convoluting by the pulse duration, we obtain a decay curve for the scattered light at the switch-off.

The advantage of the CD model is that it includes interference and cooperative effects.
One can also include temperature effects by using time dependent positions of the atoms~\cite{Bienaime2012,Eloy2018}.
However, computing capabilities limit its use to a few thousands of atoms and it is thus hard to explore large optical depths without introducing spurious high-density effects.
The random walk model does not suffer from this limitation and can be applied with the parameters of our experiment.
It can also easily account for some experimental imperfections, like the finite linewidth of the laser spectrum.
Doppler broadening can also be included ``by hand'' by a probabilistic frequency shift at each scattering~\cite{Eloy2018}, also accounting for subtle effects like the correlation between the frequency shift and the initial detuning and the scattering angle (see, e.g.,~\cite{Carvalho2015}).
However, all coherent and interference effects are neglected. Therefore, comparing the results given by the two models helps identify the relevant physics.

\subsection{Comparison between the coupled-dipole and the random walk models in the ideal case}\label{sec:numT0}

In this section we consider motionless atoms ($T=0$).
In the CD equations, the driving beam profile $\Omega(\bm{r})$ is a truncated plane wave of radius $R/2$, where $R$ is the rms radius of the atomic cloud.
In the RW simulations, the excitation beam is infinitely narrow and centered on the cloud. In the two models the driving field is perfectly monochromatic.

\begin{figure}
\centerline{\includegraphics{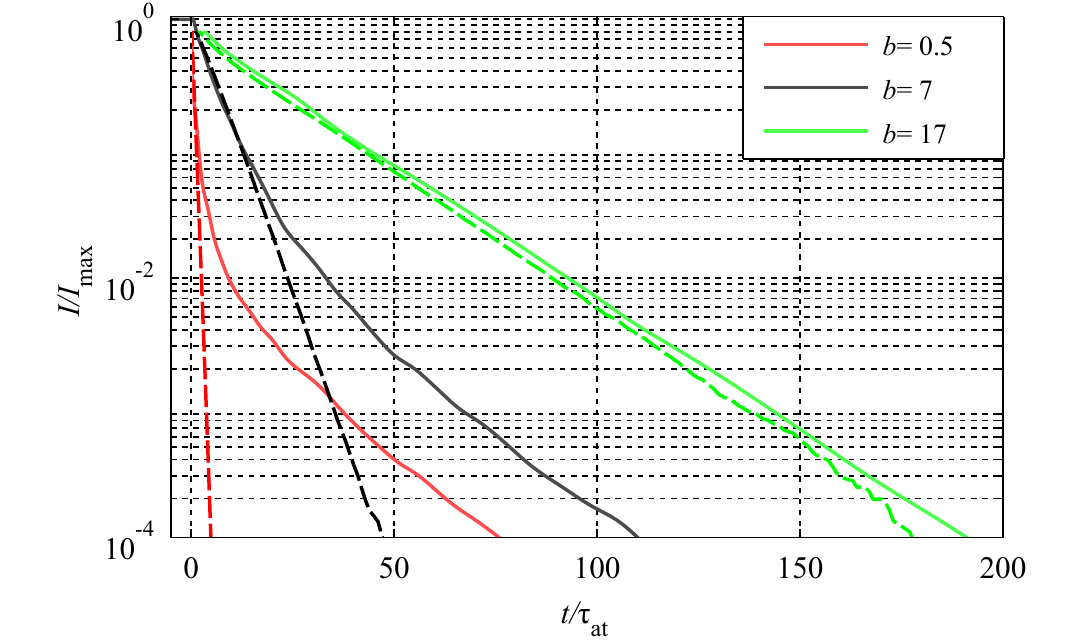}}
\caption{Numerical simulations of the decay for a fixed $b_0=17$ and different detunings $\delta = \{0, 0.6, 2.9\}$ in order to vary the optical depth (legend). The solid lines represent the calculations for the coupled-dipole model, the dashed lines show the result for the random walk model. The two models are in agreement at high $b$. For large $b_0$ and moderate $b$ (slightly detuned excitation), radiation trapping dominates the decay at the beginning and subradiance dominates at the end. For very large detuning and very low $b$, superradiance at early times would be visible in the CD model~\cite{Araujo2016}.}
\label{fig:decay_theo}
\end{figure}
Examples of decay curves for different optical depths $b$ are shown in figure~\ref{fig:decay_theo}.
Solid lines are computed from the CD equations and the dashed lines from RW simulations.
Here, the resonant optical depth is fixed, $b_0 = 17$, and the optical depth is changed by varying the detuning.
The data for the highest $b$ corresponds to $\delta=0$.
The main observation is that the two models are in good agreement for the highest optical depth, showing that in this case, radiation trapping completely dominates the decay dynamics, and subradiance is not or hardly visible.
As the detuning increases and the optical depth decreases accordingly, while $b_0$ remains large, radiation trapping becomes less and less important.
It still dominates the early decay (superradiance is not visible above $b\sim 1$~\cite{Araujo2016}) but subradiance dominates afterwards.

\begin{figure}
\centerline{\includegraphics{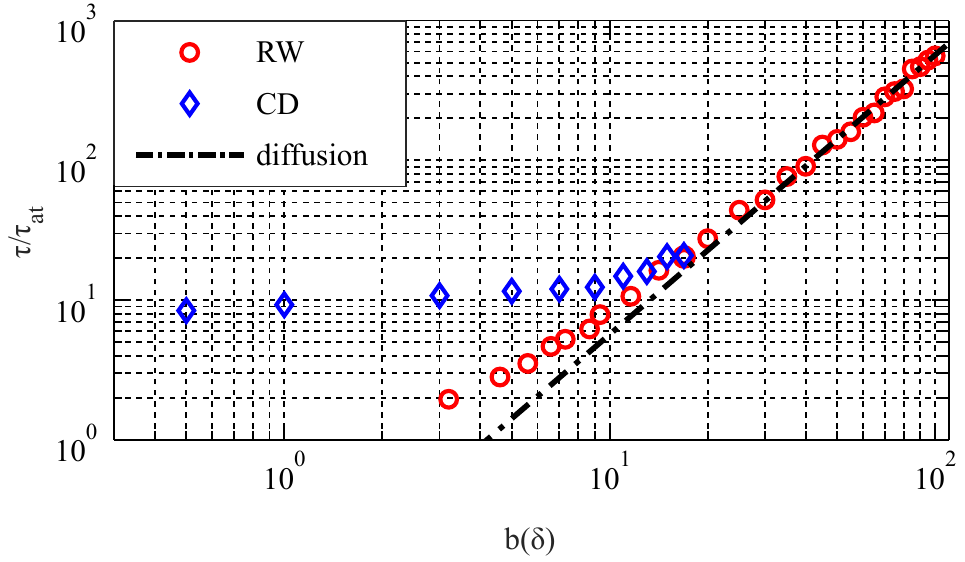}}
\caption{Comparison of the late decay time in different models. The optical depth $b(\delta)$ is changed by varying the detuning and keeping the on-resonant optical depth constant ($b_0=17$). Red circles correspond to random walk simulations, blue diamonds to the coupled-dipole model and the dash-dotted line to the diffusion model (Eq. 9).}
\label{fig:tau_theo}
\end{figure}

A systematic comparison between the two models is performed in figure~\ref{fig:tau_theo}, in which we plot the late decay time determined by an exponential fit in the amplitude range $\left[10^{-3} \; 10^{-4}\right]$.
We also show the prediction of a diffusion model for multiple scattering,
\begin{equation}
\tau_\mathrm{diff} = \frac{3 b^2}{\alpha \pi^2} \, \tau_\text{at} \, ,
\end{equation}
with $\alpha \approx 5.35$ for a Gaussian density distribution~\cite{Labeyrie2003}.

Figure~\ref{fig:tau_theo} shows that the decay computed by the RW simulation tends toward the asymptotic behavior described by the diffusion equation, which is a good approximation for optical depth larger than $b\sim 20$.
More interestingly, the CD model also starts to reach this asymptotic behavior and gives results very close to the RW model above $b\sim 10$.
On the contrary, at low $b$ (large $\delta$), the CD model levels to a constant value for the decay time, which corresponds to subradiance, not included in the RW model.

Similar comparisons (not shown here) for resonant excitation and different $b_0$ show the same behavior: the two models are in agreement above $b \sim 10$ while for smaller $b$ subradiance is visible in the CD model.

To conclude, in this idealized scenario (narrow exciting beam, $T=0$), subradiance dominates the slow switch-off dynamics for small $b$ and radiation trapping dominates for large $b$, as expected from the scaling behaviors, respectively linear in $b_0$ and quadratic in $b$.
Moreover, although the deep multiple scattering regime is hard to explore, these results confirm that radiation trapping is well included in the CD model.

\subsection{Comparison between experimental data and random walk simulations}\label{sec:num_T}

The situation is not so simple in the experiment because of a number of effects. As already discussed in~\cite{Labeyrie2003,Labeyrie2005}, the two most important effects are the temperature and the spectrum of the incident laser.
First, frequency redistribution during multiple scattering due to Doppler broadening breaks the $b^2$ scaling law, and can even make it close to a linear scaling~\cite{Pierrat2009}.
Second, the finite spectrum of the incident laser, with possibly broad wings, can be a source of resonant photons when a moderate detuning is chosen. By combining the two effects, spurious resonant photons could mimic subradiance.
Fortunately, these two effects can be included in random walk simulations, which allows us to check that the slow decay due to this spurious radiation trapping is well below the measured slow decay that we attribute to subradiance.
We have also checked that a number of other imperfections, such as a slight anisotropy of the cloud or a small misalignment of the beam from the cloud center, are indeed negligible with our parameters (see \ref{sec:imp}).

\begin{figure}
\centerline{\includegraphics[width=\columnwidth]{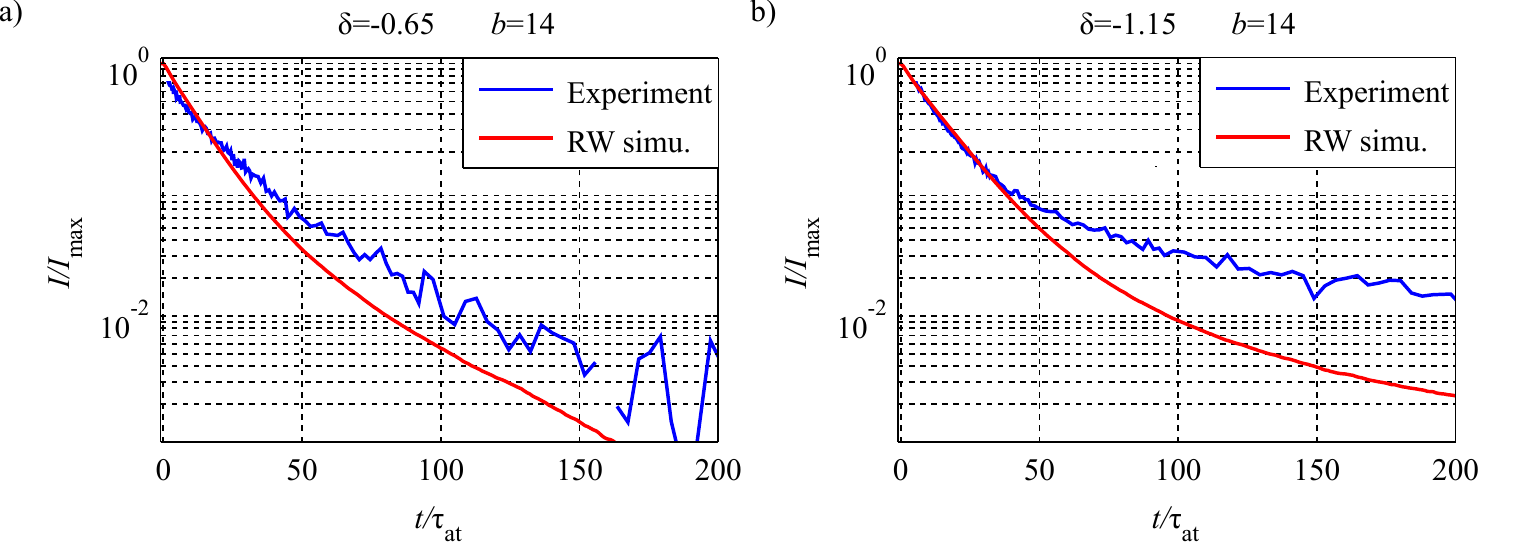}}
\caption{Direct comparison between experimental decay and simulated decay with a random walk model. The parameters of the simulation are the experimental ones. The optical depth is the same in the two panels, $b \approx 14$. (a) $b_0 = 78$, $\delta = -0.65$. (b) $b_0 = 182$, $\delta = -1.15$. The beginning of the decay is in good agreement with the RW simulation; the slower experimental decay at late time is due to subradiance.}
\label{fig:comparison}
\end{figure}

Figure~\ref{fig:comparison} shows the direct comparison between normalized experimental data and RW simulations performed with the experimental parameters, for the same $b(\delta) \approx 14$ but different $b_0$ and $\delta$.
Since the optical depths are the same, in the ideal case the two RW simulations would give the same results.
Their small difference is due to the temperature ($T=$ \SI{100}{\micro \kelvin}) and laser spectrum (FWHM = \SI{500}{\kilo \hertz}), which have different effects depending on $b_0$ and $\delta$.
The experimental data, however, have a much larger difference.
They are very close to the simulations at early time, which confirms that the measured intermediate decay is well explained by radiation trapping.
On the contrary, at long time, the experimental data are significantly above the simulations, a difference which increases with $b_0$. This is well consistent with subradiance, absent in the RW model, which dominates at long time.

\begin{figure}
\centerline{\includegraphics{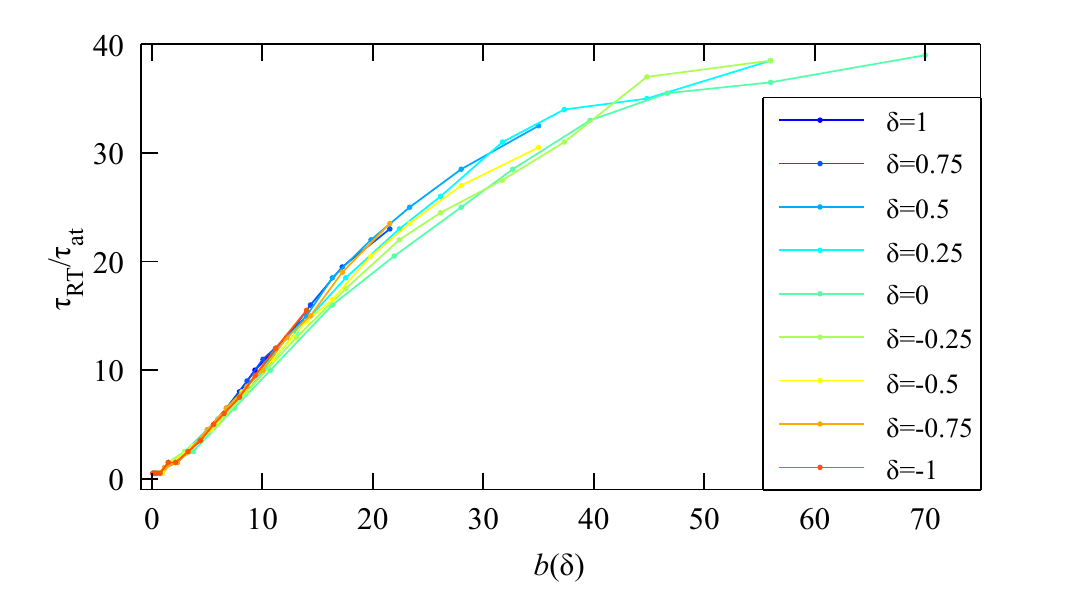}}
\caption{Numerical decay times $\tau_\text{RT}$ as a function of the optical depth $b(\delta)$ for different detunings and $b_0$. These results have to be compared with the experimental data reported in figure~\ref{fig:tau_small}(b), which shows a fair agreement between radiation trapping measurements and our random walk model.}
\label{fig:tau_theo2}
\end{figure}

Moreover, the RW simulations allow the direct comparison with the measured intermediate decay time reported in figure~\ref{fig:tau_small}(b).
Using the same definition for extracting $\tau_\text{RT}$ from the simulated decay, we report in figure~\ref{fig:tau_theo2} the results of systematic simulations for different $b_0$ and $\delta$, plotted as a function of $b(\delta)$.
As previously, the simulations are performed with the parameters of the experiment including the effects of the temperature and laser spectrum.
Therefore the decay times do not follow the quadratic behavior expected for the ideal case of zero temperature.
With these effects the decay time increases almost linearly with the optical thickness and saturates for large optical thickness.
It shows a fair agreement with the experimental data of figure~\ref{fig:tau_small}(b), without any free parameter, although we observe a discrepancy for the largest optical thickness.
Indeed above $b\approx 25$ the time $\tau_\text{RT}$ saturates faster in the experimental data than in the simulations.
This could be due to a loss process for the light during multiple scattering, for instance inelastic scattering (Raman scattering, light-induced collisions, scattering by the hot vapor background, etc.).

It is interesting to note that despite several experiments on radiation trapping in cold atoms, it is still challenging to observe a clear quadratic dependence of the radiation trapping time with the optical thickness. Indeed one needs at the same time a large cloud (such that the exciting beam can reasonably be smaller), a large optical thickness to be deep in the diffusive regime, and a very cold sample such that frequency redistribution is negligible. More precisely, one needs $bk_0v << \Gamma$, where $v$ is the rms width of the velocity distribution \cite{Labeyrie2003, Labeyrie2005, Pierrat2009}. This condition comes from the Doppler shift at each scattering event, which induces a random walk of the light frequency of step $kv$, thus producing a broadening given by $k_0v$ times the square root of the number of scattering events, i.e. $b$. Taking $b=50$ and $bk_0v = 0.1 \Gamma$ gives a temperature $T \approx$ \SI{1}{\micro K}.

\section{Summary} \label{sec:summary}

In summary, we have demonstrated that with a large cold atomic cloud of $^{87}$Rb driven by a weak laser near resonance, we can observe two different types of slow decay of the scattered light when the laser is switched off.
Moreover, with appropriate parameters, the two slow decays appear simultaneously.
At early and intermediate time, the decay is mainly due to radiation trapping, i.e. classical multiple scattering.
It is well explained by a random walk description.
At late time, subradiance creates an even slower decay.
We find that, at large enough optical depth and at zero temperature, radiation trapping could dominate the whole decay dynamics.
However, temperature-induced frequency redistribution limits radiation trapping and in our experiment, subradiance always dominates at late time.

Following previous independent observations of radiation trapping~\cite{Labeyrie2003,Labeyrie2005} and subradiance~\cite{Guerin2016} as well as a theoretical analysis of the nature of collective long-lived modes of the effective atomic Hamiltonian~\cite{Guerin2017a}, these new results significantly contribute to clarify the interplay between radiation trapping and subradiance, their dependence with experimental parameters, and more generally the physical interpretation of the slow decay at the switch-off.
This is crucial for further use of this kind of experiments for probing more subtle phenomena, as it has been proposed, for instance, for the experimental observation of Anderson localization of light in cold atoms~\cite{Skipetrov2016b}.

\section*{Acknowledgments}

The authors would like to thank Guillaume Labeyrie for fruitful discussions and Ana Cipris for help in final calibration experiments.
This work was supported by the ANR (project LOVE, Grant No. ANR-14-CE26-0032), the CAPES and the Deutsche
Forschungsgemeinschaft (WE 6356/1-1).
\appendix

\section{Transport time of light in cold atoms}\label{sec:A_tau}

The transport time is the sum of the group delay between two scattering events and the delay corresponding to the elastic scattering process, called Wigner's delay time $\tau_\text{W}$~\cite{Lagendijk1996,Labeyrie2003}:
\begin{equation}
\tau_\text{tr} = \tau_\text{W} + \frac{\ell_\text{sc}}{v_\text{g}} \, ,
\label{eq:t_tr2}
\end{equation}
where $\ell_\text{sc}$ is the mean free path and $v_\text{g}$ the group velocity.

The fundamental ingredient to compute the different terms is the atomic polarizability,
\begin{equation}
\alpha(\omega) = \frac{6\pi}{k_0^3}\times \frac{-2(\omega-\omega_0)/\Gamma + i }{1+4(\omega-\omega_0)^2/\Gamma^2} \, ,
\end{equation}
where $k_0=\omega_0/c$ is the wavevector associated to the transition.
Note that the prefactor $6\pi/k_0^3$ can also be written $\sigma_0/k_0$ with $\sigma_0$ the resonant scattering cross section. For simplicity, we will also use the notation $\mathcal{L}(\omega)$ for the Lorentzian function
\begin{equation}
\mathcal{L}(\omega) = \frac{1}{1+4(\omega-\omega_0)^2/\Gamma^2} \, .
\end{equation}

The Wigner delay time is given by the energy derivative of the dephasing acquired at the scattering~\cite{Wigner1955,Bourgain2013},
\begin{equation}
\tau_\text{W} = \frac{\partial \phi}{\partial \omega} \, .
\end{equation}
This phase is actually the argument of the polarizability,
\begin{equation}
\phi(\omega) = \arctan \left( \frac{-\Gamma/2}{\omega-\omega_0}\right) \, ,
\end{equation}
which gives
\begin{equation}
\tau_\text{W} = \frac{2}{\Gamma} \mathcal{L}(\omega) \, .
\label{eq:t_w}
\end{equation}

The mean free path is related to the scattering cross section, proportional to the imaginary part of the polarizability,
\begin{equation}
\ell_\text{sc} = \frac{1}{\rho \sigma_\text{sc}} = \frac{1}{\rho \sigma_0 \mathcal{L}(\omega)} \, .
\label{eq:l_sc}
\end{equation}

Finally, the group velocity is defined by
\begin{equation}
v_\text{g} = \frac{\partial \omega}{\partial k} \, ,
\end{equation}
with $k = n k_0$ and $n$ the refractive index.
It follows
\begin{equation}
\frac{1}{v_\text{g}} = \frac{n}{c} + k_0 \frac{\partial n}{\partial \omega} \,.
\label{eq:v_g}
\end{equation}
The refractive index is given by the real part of the polarizability,
\begin{equation}
n-1 = \frac{\rho}{2} \mathrm{Re}(\alpha) \,.
\end{equation}
Using
\begin{equation}
\mathrm{Re}(\alpha) = \frac{\sigma_0}{k_0} \times -\frac{2(\omega-\omega_0)}{\Gamma} \mathcal{L}(\omega)
\label{eq:a11}
\end{equation}
and combining equations~\ref{eq:l_sc} and ~\ref{eq:v_g} - \ref{eq:a11}, we obtain
\begin{equation}
\frac{\ell_\text{sc}}{v_\text{g}} = \frac{n\ell_\text{sc}}{c} + \frac{1}{2\mathcal{L}(\omega)} \frac{\partial}{\partial \omega}\left(-\frac{2(\omega-\omega_0)}{\Gamma}\mathcal{L}(\omega)  \right) \,.
\end{equation}
At this stage we consider that the first term is negligible, which is true for sample of reasonable size because $R \ll c/\Gamma$.
We thus obtain
\begin{equation}
\frac{\ell_\text{sc}}{v_\text{g}} = -\frac{1}{\Gamma} \left[ 1 + \frac{(\omega-\omega_0)\mathcal{L}'(\omega)}{\mathcal{L}(\omega)} \right] \,.
\end{equation}
where $\mathcal{L}'(\omega)$ is the derivative of the Lorentzian function,
\begin{equation}
\mathcal{L}'(\omega) = -\frac{8(\omega-\omega_0)}{\Gamma^2}\mathcal{L}(\omega)^2 \,.
\end{equation}
We obtain
\begin{equation}
\frac{\ell_\text{sc}}{v_\text{g}} = -\frac{1}{\Gamma} \left[ 1 -\frac{8(\omega-\omega_0)^2}{\Gamma^2} \mathcal{L}(\omega)\right] \,.
\label{eq:l_vg}
\end{equation}

Finally, combining equations~\ref{eq:t_w} and \ref{eq:l_vg} in equation~\ref{eq:t_tr2} leads to
\begin{eqnarray}
\tau_\text{tr}  &= \frac{1}{\Gamma} \left[ 2 \mathcal{L}(\omega) - 1 + \frac{8(\omega-\omega_0)^2}{\Gamma^2} \mathcal{L}(\omega)\right] \\
&= \frac{1}{\Gamma}  \left\{ 2 \mathcal{L}(\omega) \left[1 + \frac{4(\omega-\omega_0)^2}{\Gamma^2}\right] -1 \right\} \\
&= \tau_\text{at} \; .
\label{eq:tau_tr}
\end{eqnarray}

Although this result is well known, writing down its derivation allows one to notice that near resonance, the Wigner time is actually \emph{larger} than the natural lifetime of the excited state, and this is compensated by a \emph{negative} group velocity.
This shows that the simple physical picture of photons bouncing between atoms with a well ordered sequence of events, with some duration for the scattering process and some duration for the propagation between atoms, is clearly a bad picture.
Yet, it works surprisingly well in a great number of circumstances.

Another remark is that random walk simulations are considered to neglect coherent and wave effects, which is true for diffraction, interference or cooperativity.
But, as far as the temporal dynamics is concerned, a bit of wave physics enters with the use of equation~\ref{eq:tau_tr}, since it relies on wave quantities like the group velocity or the dephasing at the scattering process.
Moreover it also relies on the refractive index, which is a coherent and collective quantity.
In this respect, the random walk model corresponds to a hybrid approach.

\section{Influence of imperfections}\label{sec:imp}

To evaluate which experimental imperfections influence the radiation trapping decay we performed a systematic study by adding one effect after the other in the random walk simulations.
We also want to check that none of these imperfections is strong enough to create spurious photons on resonance, which would mimic subradiance.
We thus used the parameters of the data taken off resonance, at $\delta=-1.15$.

\begin{figure}[h]
\centerline{\includegraphics{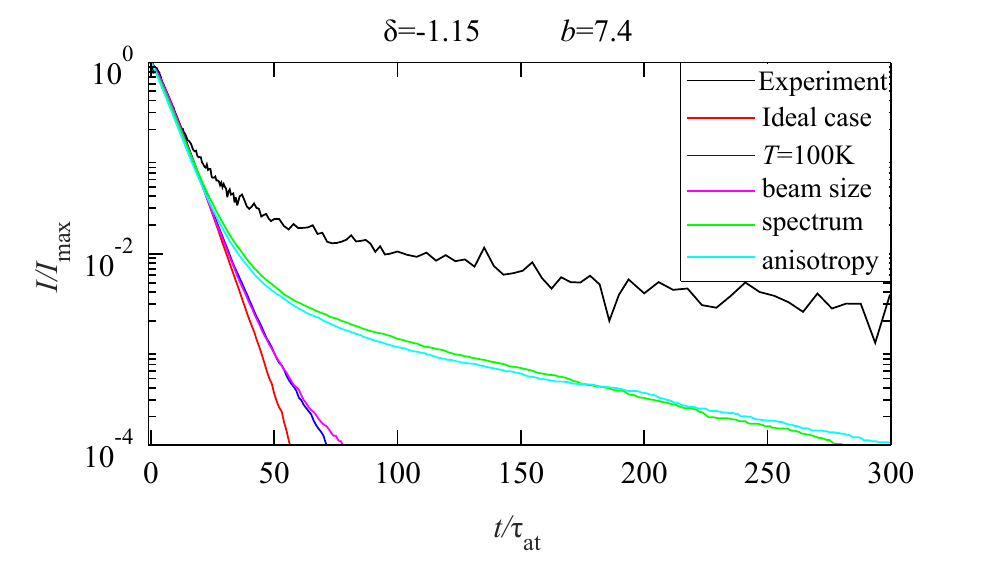}}
\caption{Comparison between experimental data and the random walk simulations for a detuning of $\delta = -1.15$ and $b = 7.4$. To show the impact of the different imperfections we included them one after the other in the random walk model. In the ideal case the temperature is zero, the probe beam is centered and infinitely small and its spectrum is perfectly monochromatic. The main effects are the finite temperature as well as the width of the laser spectrum.}
\label{fig:imper}
\end{figure}

The curves are shown in figure \ref{fig:imper}, starting from the ideal case of zero temperature, an infinitely narrow beam crossing the sample at its center, a spherical Gaussian atomic distribution and a perfectly monochromatic light.
First the finite temperature is added, which changes the slope of the late decay.
Afterward the finite beam size is added, which does not affect the decay much since the beam is still much smaller than the cloud.
Next the finite width of the probe spectrum ($FWHM=$ \SI{500}{\kilo \hertz}), including its measured Lorentzian wings, is added, which leads to a significant change in the decay time as well as in the relative level of the slow decay.
This is due to the resonant photons contained in the broad Lorentzian wings of the laser spectrum.
Finally the slight anisotropy in the cloud shape is added, which only leads to a minor change.
We also display the corresponding experimental data, which shows that the slow decay at late time, attributed to subradiance, is indeed well above the simulated radiation trapping decay.

\newpage

\section*{References}
\bibliographystyle{unsrt}
%\bibliography{Literatur_Nice}

\end{document}